\documentclass[12pt]{article}
\usepackage{indentfirst} 
\usepackage{amsfonts,amssymb,amsmath}
\usepackage{bm}          
\usepackage{cite}
\usepackage{url}
\usepackage{enumitem}
\usepackage[usenames,dvipsnames]{xcolor}
\usepackage{todonotes}
\usepackage{multicol}
\usepackage{MnSymbol}
\usepackage{hyperref}
\usepackage{tensor}
\usepackage{arcs}
\usepackage[titletoc,title]{appendix}

\usetikzlibrary{arrows}

\AtBeginDocument{}%

\newcommand{\corurl}{red}
\newcommand{\corcite}{ForestGreen}
\newcommand{\corlink}{blue}

\newcommand{\dd}{\mathrm{d\!l}}

\newcommand{\st}{\ \Big\rvert \ }

\hypersetup{linktocpage,colorlinks,urlcolor=\corurl,citecolor=\corcite,linkcolor=\corlink,
pdftitle={Free scalar field theory on a Sobolev space over a bounded interval},
pdfauthor={J.F. Barbero G.}}

\numberwithin{equation}{section}  

\def\QED{{\boldmath$\rule{0.5em}{0.5em}$}}                                
\def\markatright#1{\leavevmode\unskip\nobreak\quad\hspace*{\fill}{#1}}    
\def\qed{\markatright{\QED}}                                              

\usepackage{titling}
\usepackage{authblk}
\title{Free scalar field theory on a Sobolev space over a bounded interval}
\author[1]{J. Fernando Barbero G.}
\affil[1]{Instituto de Estructura de la Materia, IEM-CSIC. Serrano 123, 28006 Madrid, Spain}
\date{}                     
\begin{document}
	
\maketitle
\date{January 30, 2023}


\begin{abstract}
This paper discusses several functional analytic issues relevant for field theories in the context of the Hamiltonian formulation for a free, massless, scalar field defined on a closed interval of the real line. The fields that we use belong to a Sobolev space with a scalar product. As we show this choice is useful because it leads to an explicit representation of the points in the fibers of the phase space (the cotangent bundle of the configuration space). The dynamical role of the boundary of the spatial manifold where the fields are defined is analyzed.
\end{abstract}

\tableofcontents


%
%
\section{Introduction}{\label{sec_intro}}

One traditional avenue for the quantization of mechanical systems and field theories relies on their Hamiltonian formulation. This was, for instance, the way in which Loop Quantum Gravity originally tackled the problem of finding a consistent quantization for General Relativity \cite{Ashtekar} though, at the present moment, alternative covariant methods (spin foams) also exist \cite{Rovelli}.

Many of the relevant field theories in physics are described by singular Lagrangians which demand a careful analysis and must be dealt with by appropriate means. Among the approaches developed for this purpose we would like to highlight the one proposed by Dirac \cite{Dirac} and the more geometrically flavored ideas put forward by Gotay, Nester and Hinds \cite{GNH1,GNH2,Gotay_thesis}. A point that is usually neglected, in particular when using Dirac's method, is the necessity to pay attention to functional analytic details associated with the precise definition of the spaces where the fields live. One of the goals of the present paper is to illustrate this point with a particular example: a free, massless, scalar field in $1+1$ dimensions defined on a finite interval of the real line.

The Lagrangian formulation of the dynamics of a mechanical system can be described and understood in geometric terms. The starting point is a finite dimensional differentiable manifold---the configuration space $Q$---and a one to one map from $Q$ to the set of  static configurations of the system. The dynamics is then introduced in a few steps. First a real function $L:TQ\rightarrow \mathbb{R}$, called the \textit{Lagrangian}, is defined in the tangent bundle $TQ$ of the configuration space. Then a set of suitably regular paths $\mathcal{P}$ in $Q$ is chosen. They are parametrized by time $t\in[t_1,t_2]$ and connect two given points $q_1,q_2\in Q$ in such a way that $q(t_1)=q_1$ and $q(t_2)=q_2$. Finally, the \textit{action} $S:\mathcal{P}\rightarrow\mathbb{R}$ is built as an integral in time of the Lagrangian suitably evaluated on the elements of $\mathcal{P}$. The dynamical evolution of the system is given by the stationary points of $S$. The stationarity conditions are the Euler-Lagrange equations.

An alternative way to look at the evolution of a mechanical system is provided by the Hamiltonian formulation in the cotangent bundle $T^*Q$ of the configuration space  (known as \textit{phase space}). The elements of the fibers in $T^*Q$ are pairs $(q,\mathbf{p}_q)$ with the momenta $\mathbf{p}_q$ belonging to the dual $(T_qQ)^*$ of the tangent space $T_qQ$. Notice that the phase space is not necessarily a Cartesian product (nor isomorphic to one) as the covectors $\mathbf{p}_q$ are elements of vector spaces associated with different base points $q$. We highlight this fact by using a $q$ subindex to label the momenta $\mathbf{p}_q$. An important reason to work in the phase space is the availability of a canonical geometric structure, the symplectic 2-form $\Omega$, that can be used to define operations---such as the Poisson brackets---which are useful to understand relevant features of the dynamics and play a significant role in some approaches to quantization.

Given the Lagrangian description for the dynamics of a given mechanical system the passage to the Hamiltonian one can be carried out in a few simple steps. The first of them is the computation of the \textit{fiber derivative} associated with the Lagrangian (the \textit{definition of canonical momenta} in physics parlance). This is a fiber preserving map $F\!L:TQ\rightarrow T^*Q:(q,v_q)\mapsto (q,\mathbf{p}_q)$. With the help of the fiber derivative $F\!L$ the energy $E:TQ\rightarrow \mathbb{R}:(q,v_q)\mapsto \mathbf{p}(v_q)-L(q,v_q)$ is then defined. A most important feature of $E$ is that, for time-independent Lagrangians, it is constant in time when evaluated on solutions to the Euler-Lagrange equations. Finally, for \textit{hyperregular systems} (those for which the fiber derivative is a diffeomorphism) it is possible to define the \textit{Hamiltonian} $H:T^*Q\rightarrow \mathbb{R}$ as $H=E\circ F\!L^{-1}$. The dynamics of the system is obtained by projecting onto $Q$ the integral curves of the Hamiltonian vector fields $X\in\mathfrak{X}(T^*Q)$ obtained by solving the equation $\imath_X\Omega=\dd H$ (here $\imath$ and $\dd$ denote the interior product and the exterior derivative in $T^*Q$, respectively). Notice that in the preceding description it is crucial to have an invertible and smooth fiber derivative. When this is not the case one has a \textit{singular system}, the treatment of which requires special care. Notice, also, that the canonical form $\Omega$ must be invertible to guarantee the solvability of the equation for the Hamiltonian vector fields. When this is not the case we say that the canonical 2-form is \textit{presymplectic}. 

The previous scheme can be exported almost \textit{verbatim} to field theories. In principle the only necessary change is to replace the finite dimensional configuration manifold by some infinite dimensional space of fields endowed with the necessary mathematical structures (topology, norms, \textit{etcetera}). Although this is apparently straightforward it demands some care. Even in those instances, such as the model discussed in the present paper, where the manifold structure of the space of fields is as simple as possible (a Hilbert space), the structures used in its definition are important and must be properly taken into account.  A concrete example of the kind of problems that crop up in the passage from the Lagrangian to the Hamiltonian formulations of field theories is the characterization of the elements of the duals of the tangent spaces $T_qQ$. Whereas this is not an issue in the finite dimensional case, it is non-trivial for field theories. In some circumstances, for instance, when the configuration space is an infinite dimensional Hilbert space, dual objects are easy to characterize with the help of the Riesz-Fr\'echet theorem. However, this is not the case for more general Banach spaces or infinite dimensional manifolds of other types.

At this point the reader may wonder about the physical relevance of the preceding discussion in the believe that the subtleties described above will not affect the usual field theories. Unfortunately this is not the case; even such basic models as the massless, free, scalar field are subtle from the perspective of functional analysis and must be approached with some care, especially in situations when the spatial manifolds where the basic fields live have boundaries.

The purpose of this paper, which builds upon previous work by the author and collaborators \cite{Barbero1}, is to discuss a specific example which illuminates many of the issues mentioned above: a free, massless scalar field defined on a closed interval of the real line. In order to carry out the analysis we will choose as the configuration manifold the Sobolev space $H^1[-\ell,\ell]$ ($\ell\in \mathbb{\mathbb{R}}$). This is a Hilbert space with a scalar product defined with the help of first order weak derivatives. Its elements are regular enough to guarantee that the standard Lagrangian for a free, massless, scalar field makes mathematical sense. It is important to keep in mind that the manifold where the fields are defined has a boundary whose role in the dynamics must be appropriately taken into account.

The structure of the paper is the following. After this introduction, Section \ref{sec_Lag} is devoted to present the model studied in the paper. We carefully define the configuration space, introduce the Lagrangian, derive the field equations and discuss their meaning. Section \ref{sec_Ham} deals with its Hamiltonian formulation. Special attention is paid to the description of the elements of the fibers in phase space and the properties of the fiber derivative. We also study the canonical symplectic form and its pullback to the primary constraint hypersurface where the dynamics takes place. The last Section \ref{sec_conclusions} is devoted to our conclusions, comments and future work on the subject. We end the paper with Appendix \ref{appendix_details} where we discuss some interesting structures and constructions in the Sobolev space $H^1[-\ell,\ell]$ and Appendix \ref{appendix_computations} where we provide some computational details necessary to understand the field equations of our model. In the following we measure times and length intervals in the same units (length). The propagation velocity appearing in the wave equation is taken to be one.


%
%
\section{Lagrangian formulation for a free, massless scalar fields in the Sobolev space $H^1[-\ell,\ell]$}{\label{sec_Lag}}

The configuration space $Q$ of the model that we consider in the present paper is the Sobolev space $H^1(I)$ where $I$ is the closed interval of the real line $[-\ell,\ell]$ with half-length $\ell\in\mathbb{R}$. The precise definition and some mathematical details on this space can be found in Appendix \ref{appendix_details}. The elements of the tangent bundle $TQ$ are pairs $(q,v_q)$ with  $q\in Q$ and $v_q\in T_qQ$. Although the vector spaces $T_qQ$ associated with different points $q\in Q$ are different (and, generically, the tangent bundle of a manifold $Q$ need not be a Cartesian product), in the present case the tangent bundle $TH^1(I)$ is isomorphic to $H^1(I)\times H^1(I)$ because $H^1(I)$ is a vector space.

The Lagrangian that we use here is $L:TQ\rightarrow \mathbb{R}$ given by
\begin{equation}\label{Lag1}
L(\phi,v_\phi)=\frac{1}{2}\big(\|v_\phi\|_{L^2}^2-\|\phi'\|_{L^2}^2\big)\,,
\end{equation}
The subindex $L^2$ in the norm appearing in \eqref{Lag1} stands for $L^2(I)$ [similarly for $H^1(I)$] and the derivative of $\phi\in H^1(I)$ is denoted as $\phi'$, (or $\partial_x\phi$, in particular, we denote the $n$-th order derivatives of $\phi$ as $\partial_x^n\phi$). This is nothing but the usual Lagrangian for a free, massless, scalar field. Notice that both terms of $L$ are well defined for $\phi,v_\phi\in H^1(I)$ because both $v_\phi$ and $\phi'$ belong to $L^2(I)$. It is also important to realize that, even though we are working in $H^1(I)$, the norm that we use in the definition of $L$ \textit{is not} the $H^1(I)$ norm but, rather, the $L^2(I)$ one because we want to deviate as little as possible from the standard Lagrangian.

In order to define the action for this model we introduce a set of smooth paths
\begin{equation}\label{paths}
\mathcal{P}(\varphi_1,\varphi_2,[t_1,t_2]):=\big\{\varphi\in C^\infty([t_1,t_2],H^1(I)),\,\varphi(t_1)=\varphi_1,\,\varphi(t_2)=\varphi_2\big\}\,,
\end{equation}
with $t_1\,,t_2\in\mathbb{R}$ and $t_1<t_2$. Notice that, in this setting, we can use the standard concept of differentiability in Banach spaces. The action for the model is a real function $S:\mathcal{P}(\varphi_1,\varphi_2,[t_1,t_2])\rightarrow \mathbb{R}$ defined for $\varphi\in\mathcal{P}(\varphi_1,\varphi_2,[t_1,t_2])$ by
\begin{equation}\label{action1}
S(\varphi)=\int_{t_1}^{t_2} L\big(\varphi(t),\dot{\varphi}(t)\big)\,\mathrm{d}t=\frac{1}{2}\int_{t_1}^{t_2}\big(\|\dot{\varphi}(t)\|_{L^2}^2-\|\varphi'(t)\|_{L^2}^2\big)\,\mathrm{d}t\,,
\end{equation}
(here the time derivative of $\varphi$ is denoted as $\dot{\varphi}$, and $\varphi'(t):=\varphi(t)'$). In order to get the field equations we must compute the variations of $S$ about a specific path $\Phi\in \mathcal{P}(\varphi_1,\varphi_2,[t_1,t_2])$, i.e. a trajectory in the field space $H^1(I)$ connecting a configuration $\varphi_1$ at $t_1$ with another one $\varphi_2$ at $t_2$. To this end it is convenient to introduce the space of variations $\delta \mathcal{P}(t_1,t_2)$ defined as
\[
\delta\mathcal{P}(t_1,t_2):=\big\{\delta\varphi\in C^\infty([t_1,t_2],H^1(I)),\,\delta\varphi(t_1)=0,\,\delta\varphi(t_2)=0\big\}\,.
\]
Notice that, as $H^1(I)$ is a vector space, for every $\lambda\in\mathbb{R}$, $\varphi\in \mathcal{P}(\varphi_1,\varphi_2,[t_1,t_2])$ and $\delta\varphi\in \delta\mathcal{P}(t_1,t_2)$ we have $\varphi+\lambda \delta\varphi\in\mathcal{P}(\varphi_1,\varphi_2,[t_1,t_2])$.

The variation of a real function $F:\mathcal{P}(\varphi_1,\varphi_2,[t_1,t_2])\rightarrow \mathbb{R}$ at $\Phi\in \mathcal{P}(\varphi_1,\varphi_2,[t_1,t_2])$ in the direction $\delta\Phi\in \delta\mathcal{P}(t_1,t_2)$ is defined as
\[
\delta_\Phi F:=\left.\frac{\mathrm{d}F(\Phi+\lambda\delta\Phi)}{\mathrm{d}\lambda}\right|_{\lambda=0}\,.
\]
In the case of the action $S$ this gives
\begin{equation}\label{var_action_1}
\delta_\Phi S=\int_{t_1}^{t_2}\left(\langle\dot{\delta\Phi}(t),\dot{\Phi}(t)\rangle_{L^2}-\langle\delta\Phi'(t),\Phi'(t)\rangle_{L^2}\right)\,\mathrm{d}t\,.
\end{equation}
Now, according to Hamilton's principle, $\Phi$ describes the evolution from $\varphi_1$ at $t_1$ to $\varphi_2$ at $t_2$ if and only if it is an stationary point of $S$, that is, if $\Phi$ is such that $\delta_\Phi S=0$ for any $\delta\Phi\in \delta\mathcal{P}(t_1,t_2)$. The usual way to proceed at this point is to factor out $\delta\Phi(t)$ by performing an integration by parts in $t$ in the first term and an integration by parts in the spatial variable in the second term. There are, however, a couple of important comments to make at this point. First, generic configurations for the model that we study here \textit{are not} twice differentiable, so we cannot blithely integrate by parts in the second term of \eqref{var_action_1}. The second is that, by proceeding in the standard way with this action, natural boundary conditions of the Neumann type appear. They come from the boundary terms in the spatial integration by parts mentioned before. In the present case it is not clear how these boundary conditions can appear because derivatives at specific points of $I$ need not be defined for objects in $H^1(I)$ despite the fact that the elements of $H^1(I)$ can be evaluated in a precise sense at any point of $H^1(I)$ (see \cite{Brezis} and also the discussion in \cite{Barbero1}).

In order to avoid the difficulties mentioned in the preceding paragraph we proceed as follows. First we rewrite \eqref{var_action_1} in the form
\begin{equation}\label{var_action_2}
\delta_\Phi S=\int_{t_1}^{t_2}\left(-\langle\delta\Phi(t),\ddot{\Phi}(t)\rangle_{L^2}-\frac{1}{a^2}\langle\delta\Phi(t),\Phi(t)\rangle_{H^1}+\frac{1}{a^2}\langle\delta\Phi(t),\Phi(t)\rangle_{L^2}\right)\,\mathrm{d}t\,,
\end{equation}
by using the definition of the scalar product in $H^1(I)$ and integrating by parts in $t$ [which is allowed because our paths in field space are smooth in the $t$ variable according to \eqref{paths}]. The next step is to rewrite the terms involving scalar products in $L^2$ as scalar products in $H^1$. The way to do it is explained in Appendix \ref{appendix_details}. The result is
\begin{equation}\label{var_action_3}
\delta_\Phi S=\int_{t_1}^{t_2}\langle \delta\Phi(t),T\left[\frac{1}{a^2}\Phi(t)-\ddot{\Phi}(t)\right]-\frac{1}{a^2}\Phi(t)\rangle_{H^1}\,\mathrm{d}t\,,
\end{equation}
where, for $v\in H^1(I)$, $T[v]$ denotes the Riesz-Fr\'echet representative of the continuous, linear functional $F_v:H^1(I)\rightarrow \mathbb{R}=\varphi\mapsto \langle v,\varphi\rangle_{L^2}$, i.e. an element $T[v]\in H^1(I)$ such that $\langle v,\varphi\rangle_{L^2}=\langle T[v],\varphi\rangle_{H^1}$ for all $\varphi\in H^1$ [the concrete form of $T$ is \eqref{def_T}]. Finally, in order to get the field equations we consider field variations of the form $\delta\Phi(t)=\delta\!f(t)\delta\psi$ with $\delta\!f(t)\in C^\infty[t_1,t_2]$ such that $\delta\!f(t_1)=\delta\!f(t_2)=0$ and $\delta\psi\in H^1(I)$ [so, obviously, $\delta\!f(t)\delta\psi$ is an element of $\delta\mathcal{P}(t_1,t_2)$]. We then have
\begin{equation}\label{var_action_4}
\delta_\Phi S=\int_{t_1}^{t_2}\delta\!f(t)\langle \delta\psi,T\left[\frac{1}{a^2}\Phi(t)-\ddot{\Phi}(t)\right]-\frac{1}{a^2}\Phi(t)\rangle_{H^1}\,\mathrm{d}t\,,
\end{equation}
which must vanish for any allowed $\delta\!f(t)$. The usual argument employing bump functions tells us then that, for all $t\in[t_1,t_2]$, we must have
\[
\langle \delta\psi,T\left[\frac{1}{a^2}\Phi(t)-\ddot{\Phi}(t)\right]-\frac{1}{a^2}\Phi(t)\rangle_{H^1}=0\,.
\]
This condition must also hold for every $\delta \psi\in H^1(I)$. As the orthogonal subspace to $H^1(I)$ is $\{0\}$, this is equivalent to
\begin{equation}\label{field_eqs}
T\left[\Phi(t)-a^2\ddot{\Phi}(t)\right]-\Phi(t)=0\,.
\end{equation}
for all $t\in[t_1,t_2]$. These are the field equations for the model that we are considering here. They are integro-differential equations in two variables. Although they look quite formidable, it is possible to solve them as we show in the following.

Some conditions that the solutions to \eqref{field_eqs} must satisfy are:
\begin{itemize}
\item The field equations \eqref{field_eqs} tell us that $\Phi(t)$ must be in the image of $T$ ($\mathrm{Im}T$) for all $t\in[t_1,t_2]$. This means that $\Phi(t)$ must have second order spatial derivatives in $H^1$ and satisfy $\Phi'(t)(-\ell)=\Phi'(t)(\ell)=0$ (see Appendix \ref{appendix_details}). As we can see we get the Neumann boundary conditions.
\item  According to \eqref{Tsegunda}, the equations \eqref{field_eqs} imply
\begin{equation}\label{wave_eq}
\ddot{\Phi}(t)-\Phi''(t)=0\,,\quad t\in[t_1,t_2]\,,
\end{equation}
because $\Phi(t)=T\big[\Phi(t)-a^2\ddot{\Phi}(t)\big]$ leads to $\Phi''(t)=T\big[\Phi(t)-a^2\ddot{\Phi}(t)\big]''$ and, as a consequence of \eqref{Tsegunda} and equations \eqref{field_eqs}, we have $T\big[\Phi(t)-a^2\ddot{\Phi}(t)\big]''=\ddot{\Phi}(t)$. We then conclude that all the solutions to the field equations \eqref{field_eqs} must satisfy the wave equation on $[t_1,t_2]\times[-\ell,\ell]$ and are fixed by choosing initial data on $\Phi$ and $\dot{\Phi}$ at $t=t_1$.
\item From the discussion in the previous item we see that the solutions to \eqref{field_eqs} must be found among those provided by d'Alembert's formula
\begin{equation}\label{DAlembert}
\Phi(t)(x)=\frac{1}{2}\big(F(x+t)+F(x-t)\big)+\frac{1}{2}\int_{x-t}^{x+t}G(\tau)\mathrm{d}\tau\,,
\end{equation}
(where, for simplicity, we have taken $t_1=0$). The functions $F,G:\mathbb{R}\rightarrow\mathbb{R}$ appearing in \eqref{DAlembert} are obtained from the initial data $\Phi(0)=f$ and $\dot{\Phi}(0)=g$ by extending $f\,,g:[-\ell,\ell]\rightarrow\mathbb{R}$ to the interval $[-\ell,3\ell]$, in such a way that the resulting functions are symmetric with respect to $x=\ell$, and then extending the result periodically to $\mathbb{R}$. The condition, originating in our choice of path space \eqref{paths}, that these solutions must have time derivatives of arbitrary order in $H^1(I)$ implies that the functions $F$ and $G$ must be infinitely differentiable. This can only happen if the odd order lateral derivatives of $f$ and $g$ at $-\ell$ and $\ell$ are zero. Notice that, owing to the smoothness of $F$ and $G$, the solutions given by \eqref{DAlembert} are also infinitely differentiable in $x$. As a consequence, the order in which time or spatial derivatives of solutions to the wave equation are taken does not matter.
\item As the field equations \eqref{field_eqs} are integral equations one expects that they encode information about the behavior of $\Phi$ at the endpoints of the interval $[-\ell,\ell]$. In order to see this we use formulas \eqref{derivatives_odd} and \eqref{derivatives_even}. The following computation, which makes use of the fact that on smooth solutions to the wave equation we can trade pairs of time derivatives by pairs of spatial derivatives and viceversa, shows that $\partial_x^{2n-1}\Phi(t)(\ell)=0$ for all $n\in \mathbb{N}$ and $t\in[t_1,t_2]$:
\begin{align*}
\partial_x^{2n-1}\Phi(t)(\ell)&=\frac{1}{a^2\sinh\left(\frac{2\ell}{a}\right)}\left(\int_{-\ell}^{\ell} \partial_t^{2n-2}\Big(\Phi(t,\xi)-a^2\ddot{\Phi}(t,\xi)\Big)\sinh\left(\frac{\xi}{a}\right)\mathrm{d}\xi\right)\cosh\left(\frac{\ell}{a}\right)\\
&-\frac{1}{2a^2\sinh\left(\frac{\ell}{a}\right)}\int_{-\ell}^{\ell} \partial_t^{2n-2}\Big(\Phi(t,\xi)-a^2\ddot{\Phi}(t,\xi)\Big)\sinh\left(\frac{\xi}{a}\right)\mathrm{d}\xi\\
&=0\,,
\end{align*}
by using the duplication formula for $\sinh(2\ell/a)$. Completely analogous computations show that $\partial_x^{2n-1}\Phi(t)(-\ell)=0$, $\partial_x^{2n-1}\dot{\Phi}(t)(\ell)=0$ and $\partial_x^{2n-1}\dot{\Phi}(t)(-\ell)=0$, for all $n\in \mathbb{N}$ and $t\in[t_1,t_2]$. Notice that, in particular, we get Neumann boundary conditions at $-\ell\,,\ell$. All the conditions that we have just found must also hold for the initial data $\Phi(t_1)$ and $\dot{\Phi}(t_1)$. No more conditions of this type appear for higher order time derivatives because, with the help of the wave equation \eqref{wave_eq}, we can write them as derivatives in the spatial variable of either $\Phi$ or $\dot{\Phi}$.
\item If one tries to repeat the previous argument for even-order derivatives in the spatial variable no new conditions come up; only trivial identities.
\end{itemize}
In summary, all the solutions to the field equations \eqref{field_eqs} must be smooth solutions to the wave equation \eqref{wave_eq} such that, for every $t\in[t_1,t_2]$ the odd order spatial derivatives of $\Phi(t)$ and $\dot{\Phi}(t)$ vanish at the boundary $\{-\ell,\ell\}$ for all $t\in[t_1,t_2]$.

Conversely, all the smooth solutions to the wave equation in $1+1$ dimensions satisfying the boundary conditions just found, are solutions to the field equations \eqref{field_eqs}. In order to show this it suffices to plug \eqref{DAlembert} into \eqref{field_eqs} and take into account the properties of the extended functions $F$ and $G$. Some hints and details about how this computation proceeds appear in Appendix \ref{appendix_computations}.

\bigskip

An alternative way to look at the dynamics of the system, which highlights some non-trivial aspects of it, is the following. Let us introduce a particular system of coordinates in the configuration space by making use of the orthonormal basis described in Appendix \ref{appendix_details}. These coordinates (``Fourier modes'') are defined by a map $\Psi: H^1(I)\rightarrow \ell^2(\mathbb{R})\times  \ell^2(\mathbb{R})$ where here $\ell ^2(\mathbb{R})$ is the Hilbert space of real sequences $\{a_k\}_{k=0}^\infty$ satisfying the condition $\sum_{k=0}^\infty a_k^2<\infty$, and endowed with the scalar product
\[
\langle a, b\rangle_{\ell^2}:=\sum_{k=0}^\infty a_k b_k\,,\quad \{a_k\}_{k=0}^\infty\,,\{b_k\}_{k=0}^\infty\in \ell^2(\mathbb{R})\,.
\]
The image of $f\in H^1(I)$ by $\Psi$ is
\[
\Psi(f)=(\{\langle f, s_k\rangle_{H^1}\}_{k=0}^\infty, \{\langle f, c_k\rangle_{H^1}\}_{k=0}^\infty)\,,
\]
with $s_k\,,c_k\in H^1(I)$ given by \eqref{coefs}.

We rewrite now the action $S$ in terms of these modes. This is straightforward (for instance, a curve in $H^1(I)$ will define a curve in $\ell^2(\mathbb{R})\times  \ell^2(\mathbb{R})$ by composition in the obvious way) so we omit the details. If we expand the field $\varphi(t)$ as
\begin{align*}
\varphi(t)(x)&=\frac{\sinh(x/a)}{\sqrt{a\sinh(2\ell/a)}}a_0(t)+\frac{1}{\sqrt{2\ell}}b_0(t)\\
             &+\sum_{k=1}^\infty\sqrt{\frac{\ell}{\ell^2+\pi^2a^2k^2}}\left[\sin\left(\frac{\pi k x}{\ell}\right)a_k(t)+\cos\left(\frac{\pi k x}{\ell}\right)b_k(t)\right]
\end{align*}
the action becomes
\begin{align*}
S(a_k,b_k)=\frac{1}{2}\int_{t_1}^{t_2}&\left[\left(\frac{1}{2}-\frac{\ell}{a\sinh(2\ell/a)}\right)\dot{a}_0^2(t)-\frac{1}{a^2}\left(\frac{1}{2}+\frac{\ell}{a\sinh(2\ell a)}\right)a_0^2(t)\right.\\
            &\hspace*{-18pt}-2\pi a_0(t)\sqrt{\frac{2}{a}\tanh\left(\frac{\ell}{a}\right)}\cdot\sum_{k=1}^\infty(-1)^kk\left(\frac{\ell}{\ell^2+a^2k^2\pi^2}\right)^{3/2}a_k(t)+\dot{b}_0^2(t)\\
            &\hspace*{-18pt}\left.+\frac{1}{\ell^2+\pi^2a^2k^2}\left(\sum_{k=1}^\infty \left(\ell^2\dot{a}_k^2(t)-\pi^2k^2 a_k(t)^2+\ell^2\dot{b}_k^2(t)-\pi^2k^2 b_k(t)^2\right)\right)\right]\,\mathrm{d}t\,.
\end{align*}
From this action it is straightforward to get the equations of motion (it just represents an infinite number of coupled oscillators). They are
\begin{align*}
&\left(\frac{1}{2}-\frac{\ell}{a\sinh(2\ell/a)}\right)\ddot{a}_0+\frac{1}{a^2}\left(\frac{1}{2}+\frac{\ell}{a\sinh(2\ell/a)}\right)a_0\\
&\hspace*{4.8cm}+\pi\sqrt{\frac{2}{a}\tanh\left(\frac{\ell}{a}\right)}\cdot\sum_{k=1}^\infty(-1)^kk\left(\frac{\ell}{\ell^2+a^2k^2\pi^2}\right)^{3/2}a_k=0\,,\\
&\ddot{b}_0=0\,,\\
&\ddot{a}_k+\frac{\pi^2k^2}{\ell^2}a_k+\pi (-1)^k\frac{k}{\ell} \sqrt{\frac{2}{a}\tanh\left(\frac{\ell}{a}\right)}\left(\frac{\ell}{\ell^2+a^2k^2\pi^2}\right)^{1/2}a_0=0\,,\\
&\ddot{b}_k+\frac{\pi^2k^2}{\ell^2}b_k=0\,.
\end{align*}
Notice that the action is not a diagonal quadratic form in the $a_k$, hence these coefficients do not represent normal modes in the standard sense. An interesting exercise would be to get the same details on the dynamics of the system presented in the main body of this section from this perspective.

%
%
\section{Hamiltonian formulation}{\label{sec_Ham}}

We study here the Hamiltonian formulation for the scalar field theory that we are considering in the paper. The arguments presented in this section nicely complement those of \cite{Barbero2} (which applied to the wave equation and the Maxwell equations in higher dimensional spaces). Before we start we remind the readers of the fact that the phase space of our model is isomorphic to $H^1(I)\times H^1(I)^*$ and, according to the Riesz-Fr\'echet representation theorem, it is also isomorphic to $H^1(I)\times H^1(I)$. This allows us to interpret the points in the phase space as pairs of the form $(\phi,\widehat{\pi}_\phi)$ with $\phi\,,\widehat{\pi}_\phi\in H^1(I)$, where $\widehat{\pi}_\phi$ stands here for the  Riesz-Fr\'echet representative of the covector $\bm{\pi}_\phi$ [i.e. $\bm{\pi}_\phi(w)=\langle \widehat{\pi}_\phi,w\rangle_{H^1}$, $\forall w\in H^1(I)$].

An important element in the Hamiltonian formulation is the canonical 2-form $\Omega$. A useful way to describe it is through its action on pairs of vector fields $\mathbb{X}\,,\mathbb{Y}\in\mathfrak{X}(T^*\!H(I))$. In the phase space that we consider here a vector field $\mathbb{X}$ has the form $((\phi,\bm{\pi}_\phi);(X_\phi, \bm{\mathrm{X}}_\pi))$ with $\phi\,,X_\phi\in H^1(I)$ and $\bm{\pi}_\phi\,,\bm{\mathrm{X}}_\pi\in H^1(I)^*$. This means that we associate  a pair $(X_\phi, \bm{\mathrm{X}}_\pi)$ with each and every point $(\phi,\bm{\pi}_\phi)$ of $T^*\!H^1(I)$.

Taking this into account, the canonical 2-form acting on $\mathbb{X}\,,\mathbb{Y}$ gives the following real function in phase space:
\begin{equation}\label{Omega}
\Omega(\mathbb{X},\mathbb{Y})=\bm{\mathrm{Y}}_\pi(X_\phi)-\bm{\mathrm{X}}_\pi(Y_\phi)\,.
\end{equation}
If we introduce the Riesz-Fr\'echet representatives for $\bm{\mathrm{X}}_\pi$ and $\bm{\mathrm{Y}}_\pi$ ($\widehat{\mathrm{X}}_{\pi}$ and $\widehat{\mathrm{Y}}_{\pi}$, respectively) the preceding expression becomes
\begin{equation}\label{Omega2}
\Omega(\mathbb{X},\mathbb{Y})=\langle\widehat{\mathrm{Y}}_{\pi},X_\phi\rangle_{H^1}-\langle\widehat{\mathrm{X}}_{\pi},Y_\phi\rangle_{H^1}\,.
\end{equation}

The first step in order to find the Hamiltonian formulation corresponding to the Lagrangian $L$ introduced in \eqref{Lag1} is to compute its associated fiber derivative. This is the fiber-preserving map $F\!L:T\!H^1(I)\rightarrow T^*\!H^1(I):(\phi,v_\phi)\mapsto (\phi,\bm{\pi}_\phi)$ with $\phi\in H^1(I)$ and $\bm{\pi}_\phi\in T_\phi^*H^1(I)$ defined by
\[
\bm{\pi}_\phi(w_\phi):=\left.\frac{\mathrm{d\,\,}}{\mathrm{d}t}L(\phi,v_\phi+tw_\phi)\right|_{t=0}\,,\quad v_\phi,w_\phi\in T_\phi H^1(I)\cong H^1(I)\,.
\]
In the physics parlance the fiber derivative is often referred to (and understood) as ``the definition of momenta'', but thinking about it as a map is often helpful.

In the model that we are considering here we have
\[
\bm{\pi}_\phi(w_\phi)=\frac{1}{2}\left.\frac{\mathrm{d\,\,}}{\mathrm{d}t}\|v_\phi+t w_\phi\|^2_{L^2}\right|_{t=0}=\langle v_\phi,w_\phi\rangle_{L^2}\,,
\]
and, hence,
\begin{equation}\label{momentum}
\bm{\pi}_\phi(\cdot)=\langle v_\phi,\cdot\rangle_{L^2}=\langle T[v_\phi],\cdot\rangle_{H^1}\Rightarrow \widehat{\pi}_\phi=T[v_\phi]\,.
\end{equation}

An important observation to make at this point is that $F\!L$ \textit{is not a bijection} from $T\!H^1(I)$ to the phase space $T^*\!H^1(I)$ because, although $T$ is injective, it is not onto (see Appendix \ref{appendix_details}). This tells us that we are dealing with a singular system. The image of $F\!L$ is the so called \textit{primary constraint surface} $\mathsf{M}_0$. For many mechanical systems and field theories, this is a submanifold and can be naturally described as the zero set of some appropriately smooth functions in phase space (the so called \textit{primary constraints}). Here we do not have any such representation; in fact, the image of $F\!L$ is dense in $H^1(I)\times H^1(I)^*$, hence no constraint function can exist because any continuous function from $H^1(I)\times H^1(I)^*$ to $\mathbb{R}$ vanishing on $F\!L(H^1(I)\times H^1(I))$ would be the zero function. Whenever this happens the Dirac method cannot be used. Notice, however, that we have an explicit parametrization of $\mathsf{M}_0$ provided by \eqref{def_T} that tells us that the primary constraint surface consists of the elements of $T^*\!H^1(I)$ such that the Riesz-Fr\'echet representatives of the covectors in the fibers are $\pi_\phi=T[v_\phi]$ with the ``parameter'' $v_\phi\in H^1(I)$. As the GNH method relies on the pullback of the relevant geometric structures in phase space onto the primary constraint submanifold $\mathsf{M}_0$, and we have an explicit parametrization thereof, we can readily use it to describe the Hamiltonian formulation for our model. Notice that the parameter space \textit{is just the tangent bundle of the configuration space} $T\!H^1(I)$. We follow this approach in the following and, just for the sake of recovering the standard formulas, we will rename the elements of $T\!H^1(I)$ as $(\phi,p_\phi)$.

The next ingredient that we need in order to get the Hamiltonian formulation by following the GNH approach is the Hamiltonian. For hyperregular systems this function is given by $H:=E\circ F\!L^{-1}$, where the energy $E$ is
\[
E:T\!H^1(I)\rightarrow \mathbb{R}:(\phi,v_\phi)\mapsto \bm{\pi}_\phi(v_\phi)-L\,.
\]
For singular systems the definition $H:=E\circ F\!L^{-1}$ may not work because $F\!L$ is either not a bijection or lacks regularity. In the present case $F\!L$ is not onto, however, as it is injective, we actually have a continuous and invertible map (with discontinuous inverse!) from $T\!H^1(I)$ to $\mathsf{M}_0$, so we can define the Hamiltonian there as $E\circ F\!L^{-1}$. The energy for the Lagrangian $L$ is
\[
E(\phi,v_\phi)=\frac{1}{2}\left(\|v_\phi\|^2_{L^2}+\|\phi'\|^2_{L^2}\right)=\frac{1}{2}\left(\langle T[v_\phi],v_\phi\rangle_{H^1}+\frac{1}{a^2}\langle\phi-T[\phi],\phi\rangle_{H^1}\right)\,,
\]
and interpreting the fiber derivative as a parametrization of $\mathsf{M}_0$, the preceding expression is also the Hamiltonian $H$,
\[
H(\phi,p_\phi)=\frac{1}{2}\left(\|p_\phi\|^2_{L^2}+\|\phi'\|^2_{L^2}\right)=\frac{1}{2}\left(\langle T[p_\phi],p_\phi\rangle_{H^1}+\frac{1}{a^2}\langle\phi-T[\phi],\phi\rangle_{H^1}\right)\,,
\]
Vector fields $\widetilde{\mathbb{X}}\in \mathfrak{X}(T\!H^1(I))$ have the form $\big((\phi,p_\phi);(X_\phi,X_p)\big)$ with $\phi,p_\phi,X_\phi,X_p\in H^1(I)$, so we have
\begin{equation}\label{dH}
\dd H(\widetilde{\mathbb{Y}})=\langle p_\phi,Y_p\rangle_{L^2}+\langle \phi',Y'_\phi\rangle_{L^2}=\langle T[p_\phi],Y_p\rangle_{H^1}+\frac{1}{a^2}\langle\phi-T[\phi],Y_\phi\rangle_{H^1}\,,
\end{equation}
where, as we did in Section \ref{sec_Lag}, we have bypassed a non-allowed integration by parts by using
\[
\langle \phi',Y'_\phi\rangle_{L^2}=\frac{1}{a^2}\big(\langle\phi,Y_\phi\rangle_{H^1}-\langle\phi,Y_\phi\rangle_{L^2}\big)=\frac{1}{a^2}\big(\langle\phi,Y_\phi\rangle_{H^1}-\langle T[\phi],Y_\phi\rangle_{H^1}\big)\,.
\]
Next we have to pullback the 2-form $\Omega$ onto $\mathsf{M}_0$. Again, we use the parametrization provided by $F\!L$ to get
\begin{align}\label{omega_pullback}
\begin{split}
\omega(\widetilde{\mathbb{X}}, \widetilde{\mathbb{Y}})&=(F\!L^*\Omega)(\widetilde{\mathbb{X}}, \widetilde{\mathbb{Y}})=\Omega\big(F\!L_*\widetilde{\mathbb{X}},F\!L_*\widetilde{\mathbb{Y}}\big)=\langle Y_p,X_\phi\rangle_{L^2}-\langle X_p,Y_\phi\rangle_{L^2}\\
&=\langle Y_p,T[X_\phi]\rangle_{H^1}-\langle Y_\phi,T[X_p]\rangle_{H^1}\,.
\end{split}
\end{align}
The third equality follows because the fiber derivative maps $(\phi,p_\phi)$ to $(\phi,\langle p_\phi,\cdot\rangle_{L^2})$ and its pushforward maps $((\phi,p_\phi);(X_\phi,X_p))$ to $((\phi,\langle p_\phi,\cdot\rangle_{L^2}); (X_\phi, \langle X_p,\cdot\rangle_{L^2}))$.
Now, the Hamiltonian vector fields $\widetilde{\mathbb{X}}$ that define the dynamics are given by the solutions to the equation $\imath_{\mathbb{\widetilde{\mathbb{X}}}}\omega=\dd H$ or, equivalently, by requiring that $\omega(\widetilde{\mathbb{X}}, \widetilde{\mathbb{Y}})=\dd H(\widetilde{\mathbb{Y}})$ for all $\widetilde{\mathbb{Y}}$. This last condition gives
\begin{align}
T[X_\phi]&=T[p_\phi]\,,\label{Ham_fields_phi}\\
T[X_p]&=\frac{1}{a^2}\big(T[\phi]-\phi\big)\,.\label{Ham_fields_v}
\end{align}
In order to continue we have to find out what extra conditions $\phi$ and $p_\phi$ must satisfy for the equations to be solvable and find $X_\phi$, $X_p$. Typically this process produces secondary constraints and tangency requirements that must be explicitly checked.

The injectivity of $T$ implies that Equation \eqref{Ham_fields_phi} is equivalent to
\begin{equation}\label{Ham_fields_phi2}
X_\phi=p_\phi\,.
\end{equation}
Equation \eqref{Ham_fields_v} can be rewritten in the form
\begin{equation}\label{Ham_fields_v2}
\phi=T[\phi-a^2 X_p]\,.
\end{equation}
This has several important consequences. First, $\phi$ must belong to the image of $T$, hence, it must have spatial derivatives up to second order in $H^1(I)$ and satisfy $\phi'(-\ell)=\phi'(\ell)=0$. Second, if we differentiate \eqref{Ham_fields_v2} twice, use \eqref{Ham_fields_v} and \eqref{Tsegunda} we get
\begin{equation}\label{Ham_fields_v3}
X_p=\phi''\,.
\end{equation}
As a consequence, equations \eqref{Ham_fields_phi} and \eqref{Ham_fields_v} imply the more familiar \eqref{Ham_fields_phi2} and \eqref{Ham_fields_v3} which, in turn, imply that the wave equation $\ddot{\phi}-\phi''=0$ must hold.

Now, as a consequence of \eqref{Ham_fields_phi2} and \eqref{Ham_fields_v3} we have the following chain of implications:
\begin{equation*}
\phi\in\mathrm{Im}T\Rightarrow X_\phi\in \mathrm{Im}T\Leftrightarrow p_\phi\in\mathrm{Im}T \Rightarrow X_p\in\mathrm{Im}T \Leftrightarrow \phi''\in\mathrm{Im}T\Rightarrow X''_\phi\in \mathrm{Im}T\Rightarrow\cdots
\end{equation*}
From them we deduce that $\partial_x^{2(n-1)}\phi\,,\partial_x^{2(n-1)}p_\phi\in\mathrm{Im}T$ for all $n\in\mathbb{N}$ so that
$\phi$ and $p_\phi$ must be smooth functions in the interval $[-\ell,\ell]$ and the following conditions must hold:
\[
\partial_x^{2n-1}\phi(-\ell)=0\,,\partial_x^{2n-1}\phi(\ell)=0\,,\partial_x^{2n-1}p_\phi(-\ell)=0\,,\partial_x^{2n-1}p_\phi(\ell)=0\,,n\in \mathbb{N}\,.
\]
We end this section with two comments. First, notice that the Lagrangian field equations \eqref{field_eqs} are a direct consequence of \eqref{Ham_fields_phi2} and \eqref{Ham_fields_v2}. Second, an argument based in d'Alembert's formula for smooth solutions to equations \eqref{Ham_fields_phi2} and \eqref{Ham_fields_v3} (and completely analogous to the one that we used in the discussion of the Lagrangian equations) shows that these  solutions are also solutions to \eqref{Ham_fields_phi} and \eqref{Ham_fields_v}.

%
%
\section{Conclusions and comments}{\label{sec_conclusions}}

In this paper we have studied the Lagrangian and Hamiltonian formulations for a free, massless, scalar field defined on a bounded interval $I$ of the real line. We have taken the Sobolev space $H^1(I)$ as our configuration space and the standard Lagrangian to define the dynamics of the system. As we have shown, the field equations---obtained by searching for the stationary points of the action---are integro-differential equations in two variables which look quite different from the standard wave equation.

By following the usual textbook treatment of this system one would expect to get the wave equation in the interior of the interval $I$ and Neumann boundary conditions from the boundary terms of the variations of the action. However, the standard manipulations in the action to get the equations of motion would only work for twice differentiable (in the spatial variable) prospective stationary points, leaving open the question of whether other less regular stationary points exist. As we have been able to show, the field equations \eqref{field_eqs} imply that their solutions are necessarily smooth solutions to the field equation in $1+1$ dimensions satisfying some conditions at the boundary. Among them, the Neumann boundary conditions. Interestingly, in addition to these, an infinite tower of conditions appear. They require the vanishing of the odd-order spatial derivatives of the field and its time derivative at the ends of the interval $I$. By looking at the solution provided by d'Alembert's formula it is easy to understand the role of these extra conditions. On the extended initial data necessary to use d'Alembert's formula for the model considered here they guarantee that the extension is smooth (which also implies that the solutions are smooth in $I$ for all $t$).

When the functional analytic details originating in our choice of configuration space are taken into account the Hamiltonian description also becomes interesting. A first observation is that the primary constraint submanifold $\mathsf{M}_0$ (the image of the tangent bundle of the configuration space $TQ$ under the fiber derivative) cannot be described as the vanishing set of some functions in phase space. This precludes the use of Dirac's method (there are no primary constraints to work with) but the GNH approach can be employed without difficulty. Actually, $TQ$ can be interpreted as a space of parameters for $\mathsf{M}_0$ so, in practice, the pullback to $\mathsf{M}_0$ used in the GNH treatment can be alternatively interpreted as a pullback to $TQ$. An interesting observation to make at this point is the fact that we have been able to deal with a field theory defined on a manifold with boundary without having to worry about the computation of Poisson brackets. This is so because we have only needed the canonical 2-form $\Omega$. This does not mean that Poisson brackets cannot be computed in the kind of functional space that we have used here. On the contrary, it is actually possible to do it (see \cite{Barbero1}) if some care is exercised. Notice, however, that the difficulties that we have mentioned regarding the possibility of representing the primary constraint surface as the zero set of smooth constraint functions somehow limit the use of the full phase space (and also the recourse to Dirac's quantization).

The analysis that we have presented here is consistent with the one in \cite{Gotay_thesis} (see, for instance, the discussion of section C. of Chapter 4). However, the final approach that we have followed is different from the one favored in that work. In \cite{Gotay_thesis} a relaxed consistency condition for the dynamics is employed (the author uses of the so called \textit{manifold domains} and a weakened definition of the meaning of tangency of vector fields). Here we have followed an approach essentially identical to the one that one would employ for a mechanical system. Arguably,  in order to illustrate some of the issues that crop up both in the Lagrangian and Hamiltonian settings we have used explicit orthonormal bases for $H^1(I)$ and, also, have made intensive use of the concrete form of the Riesz-Fr\'echet representative $T$ of the functional that gives the canonical momenta in terms of the velocities. However, the extension of the present analysis for more complicated systems (i.e. in higher dimensions, other choices of functional spaces and introducing interactions) seems to be feasible and will be the matter of future work.

Although the problems discussed in the paper can be considered as a fine print of sorts (at the end of the day we end up getting the familiar scalar field dynamics), the solutions that we have found to them provide a deeper understanding of the system which may be useful when considering its quantization or the quantization of interacting models derived from it. From the image of quantum dynamics provided by path integrals, it is to be expected that the contributions of non-stationary paths may pay a significant role. Several actions with the same classical dynamics but defined on different spaces may describe different quantum dynamics for the same classical system as the contributions of non-stationary paths may be different. Understanding how this happens in practice for extensions of the model studied here is an interesting problem that will be the subject of future work.

%
%
\section*{Acknowledgments}

The author wants to thank M. Basquens, B. D\'{\i}az,  J. Margalef-Bentabol, Valle Varo and Eduardo J. S. Villase\~nor for interesting discussions and comments, and the organizers of the EREP held in Bilbao in July, 2023 for their invitation to participate in the conference and talk about a subject related to the content of this paper. This work has been supported by the Spanish Ministerio de Ciencia Innovaci\'on y Universidades-Agencia Estatal de Investigaci\'on grant AEI/PID2020-116567GB-C22.

\begin{appendices}

%
%
\section{The Sobolev space $H^1[-\ell,\ell]$}\label{appendix_details}

The Sobolev space $H^1(I)$ with $I=[-\ell,\ell]$, $\ell\in \mathbb{R}$ is defined as (see \cite{Brezis})
\begin{equation}\label{Sobolev_1dim}
H^1(I):=\Big\{ u\in L^2(I) \st \exists g\in L^2(I) \ : \ \    \int_I u \varphi'=-\int_I g\varphi \ , \ \forall \varphi\in C_c^1(I) \Big\}\,.
\end{equation}
where $C_c^1(I)$ is the space of continuously differentiable real functions with compact support on $I$ (test functions). The elements of $H^1(I)$ belong to $L^2(I)$ and have \textit{weak derivatives}, also in $L^2(I)$, defined on test functions by integration by parts. In the following we will write $g=u'$ and sometimes denote $H^1(I)$ as $H^1[-\ell,\ell]$. In this function space we can define the scalar product of two elements $u\,,v\in H^1(I)$
\begin{equation}\label{scalar_product}
\langle u,v\rangle_{H^1}:=\int_I\big(uv+a^2u'v'\big)\,,
\end{equation}
($a>0$) and the associated norm $\|u\|^2_{H^1}:=\langle u,u\rangle_{H^1}$. Endowed with $\langle \cdot,\cdot\rangle_{H^1}$, $H^1(I)$ is a separable Hilbert space. Notice that we have introduced a length parameter $a$ to control the relative weight in the scalar product of the terms involving functions and their derivatives. Although, as shown in the paper, it does not play any significant role in the dynamics of the free scalar field, it is interesting to see how it appears in the orthonormal basis discussed below and may play some role in the quantization of the system.

For many practical purposes the following orthonormal basis of $H^1(I)$ comes in handy
\begin{align}\label{coefs}
\begin{split}
  & s_0 (x)=\frac{1}{\sqrt{a\sinh(2\ell/a)}}\sinh(x/a)\,, \\
  & c_0 (x) =\frac{1}{\sqrt{2\ell}}\,, \\
  & s_k (x)=\sqrt{\frac{\ell}{\ell^2+\pi^2 a^2 k^2}}\sin(\pi k x/\ell)\,,\quad k\in\mathbb{N}\,, \\
  & c_k (x)=\sqrt{\frac{\ell}{\ell^2+\pi^2 a^2 k^2}}\cos(\pi k x/\ell)\,,\quad k\in\mathbb{N}\,,
\end{split}
\end{align}
with $x\in[-\ell,\ell]$. In this basis an element $f$ of $H^1(I)$ can be expanded as
\begin{align*}
f(x)=&\frac{a_0}{\sqrt{a\sinh(2\ell/a)}}\sinh(x/a)+\frac{b_0}{\sqrt{2\ell}}\\
&+\sum_{k=1}^\infty\sqrt{\frac{\ell}{\ell^2+\pi^2 a^2 k^2}}\Big(a_k\sin(\pi k x/\ell)+b_k\cos(\pi k x/\ell)\Big)\,,
\end{align*}
and the square of the norm of $f\in H^1(I)$ is
\begin{align*}
\|f\|^2_{H^1}=\sum_{k=0}^\infty(a_k^2+b_k^2)<+\infty\,.
\end{align*}
It is interesting to point out that the $a_0$ coefficient is
\begin{equation}\label{a0}
a_0=\sqrt{\frac{a}{\sinh(2\ell/a)}}\big(f(\ell)-f(-\ell)\big)\cosh(\ell/a)\,,
\end{equation}
where $f(\ell)$ and $f(-\ell)$ are well defined because every $f\in H^1(I)$ is equal, almost everywhere, to a continuous function $\widetilde{f}$ defined on $[-\ell,\ell]$ (see \cite{Brezis}). The expression \eqref{a0} provides a simple characterization of the elements of $H^1(I)$ orthogonal to $s_0$, i.e. they are those satisfying $\widetilde{f}(\ell)=\widetilde{f}(-\ell)$. It is also interesting to point out that an orthonormal basis of $L^2(I)$ is given by
\begin{align*}
  & C_0 (x) =1/\sqrt{2\ell}\,, \\
  & S_k (x)=\sin(\pi k x/\ell)/\sqrt{\ell}\,,\quad k\in\mathbb{N}\,, \\
  & C_k (x)=\cos(\pi k x/\ell)/\sqrt{\ell}\,,\quad k\in\mathbb{N}\,,
\end{align*}
the elements of which are in obvious correspondence with those of the $H^1(I)$ basis given above \textit{with the notable exception of} $s_0$.

A useful result that we use in the paper is the following. Let $v\in H^1(I)$, then $F_v:H^1(I)\rightarrow \mathbb{R}=\varphi\mapsto \langle v,\varphi\rangle_{L^2}$ is a continuous linear functional because
\[
|F_v(\varphi)|=|\langle v,\varphi\rangle_{L^2}|\leq\|v\|_{L^2}\|\varphi\|_{L^2}\leq \|v\|_{L^2}\|\varphi\|_{H^1}\,.
\]
Its Riesz-Fr\'echet representative $T[v]\in H^1(I)$, satisfying $F_v(\varphi)=\langle T[v],\varphi\rangle_{H^1}$, for all $\varphi\in H^1(I)$, is
\begin{align}\label{def_T}
\begin{split}
T[v](x)&=\frac{1}{a\sinh\left(\frac{2\ell}{a}\right)}\left(\int_{-\ell}^\ell v(\xi)\sinh\left(\frac{\xi}{a}\right)\mathrm{d}\xi\right)\sinh\left(\frac{x}{a}\right)\\
                 &+\frac{1}{2a\sinh\left(\frac{\ell}{a}\right)}\,\,\int_{-\ell}^\ell v(\xi)\cosh\left(\frac{|x-\xi|-\ell}{a}\right)\mathrm{d}\xi\,.
\end{split}
\end{align}
This result can be obtained by working in the orthonormal basis introduced above, using the fact that, if we denote the basis elements as $e_k$, we have $T[v]=\sum_kF_v(e_k)e_k$, and the identity
\begin{equation}\label{identity1}
  \sum_{k\in \mathbb{Z}}\frac{e^{i\alpha k y}}{1+\alpha^2 k^2}=\frac{\pi}{\alpha\sinh\left(\frac{\pi}{\alpha}\right)}\cosh\left(|y|-\frac{\pi}{\alpha}\right)\,,
\end{equation}
valid for $y\in[-\frac{2\pi}{\alpha},\frac{2\pi}{\alpha}]$. A quick way to arrive at \eqref{identity1} is to compute the Fourier coefficients of the function $f:\mathbb{S}^1\rightarrow \mathbb{R}:e^{i\theta}\mapsto \cosh((|\theta|-\pi)/\alpha)$ with $\theta\in(-\pi,\pi]$
\[
\hat{f}_k=\frac{1}{2\pi}\int_{-\pi}^\pi e^{-i k \theta}\cosh\left(\frac{|\theta|-\pi}{\alpha}\right)\mathrm{d}\theta=\frac{\alpha}{\pi}\sinh\left(\frac{\pi}{\alpha}\right)\frac{1}{1+\alpha^2 k^2}\,,
\]
so that
\[
\frac{\alpha}{\pi}\sinh\left(\frac{\pi}{\alpha}\right)\sum_{k\in \mathbb{Z}}\frac{e^{ik\theta}}{1+\alpha^2 k^2}=\cosh\left(\frac{|\theta|-\pi}{\alpha}\right)\,,
\]
from which \eqref{identity1} immediately follows by taking $\theta=\alpha y$.

The continuous map $T:H^1(I)\rightarrow H^1(I):v\mapsto T[v]$ can be interpreted as an integral transform of sorts. The first term in \eqref{def_T} is a smooth function and the second is a convolution. A well-known fact about convolutions is their smoothing property. In the present case a straightforward calculation applying Leibniz's integral rule after splitting the second integral in \eqref{def_T} into two integrals on the intervals $[-\ell,x]$ and $[x,\ell]$, gives
\begin{equation}\label{Tsegunda}
T[v]''=\frac{1}{a^2}\big(T[v]-v\big)\,.
\end{equation}
This result has some important consequences. First, it shows that $T[v]''\in H^1(I)$ because both $T[v]$ and $v$ are in $H^1(I)$ (as a consequence, $T[v]'''\in L^2(I)$ but, generically, cannot be differentiated more than three times). Also, the map $T:H^1(I)\rightarrow H^1(I):v\mapsto T[v]$ is not onto owing to the smoothness properties of $T[v]$ just mentioned. As discussed in the main text of the paper this means that the fiber derivative will not be a diffeomorphism between the tangent bundle of the configuration space and the phase space.

Another important property of $T[v]$, which can be found by a direct computation,  is that its lateral derivatives at the ends of the interval $[-\ell,\ell]$ vanish (i.e. $T[v]'(-\ell)=0$ and $T[v]'(\ell)=0$). This fact, together with \eqref{Tsegunda}, provides us with a useful characterization of the image of $T$ as the set of solutions to the second order linear differential equation
\begin{equation}\label{eq_diff}
a^2 y''-y=-v
\end{equation}
satisfying the boundary conditions $y'(-\ell)=0$, $y'(\ell)=0$ for arbitrary $v$ in $H^1(I)$. The solutions to \eqref{eq_diff} are given by \eqref{def_T}. Notice that any function $y$ with derivatives up to third order in $L^2(I)$ satisfying the boundary conditions $y'(-\ell)=0$, $y'(\ell)=0$ will be a solution to \eqref{eq_diff} for some $v\in H^1(I)$, hence, we arrive at the final---and simplest---characterization of the image of $T$:
\[
\mathrm{Im}\,T=\{y\in H^1(I): y', y''\in H^1(I)\,\,\mathsf{and}\,\,y'(-\ell)=y'(\ell)=0\}\,.
\]

An important observation is that $T$ is injective. This can be proved by showing that if $T[v]=0$ for $v\in H^1(I)$, then $v=0$. Indeed, let us take $v\in H^1(I)$. Now
\[
T[v]=\langle v,s_0\rangle_{L^2}s_0+\langle v,c_0\rangle_{L^2}c_0+\sum_{k=1}^\infty\langle v,s_k\rangle_{L^2}s_k+\sum_{k=1}^\infty\langle v,c_k\rangle_{L^2}c_k=0
\]
implies
\begin{align}
&\langle v,s_0\rangle_{L^2}=0\,,\label{cond1}\\
&\langle v,c_0\rangle_{L^2}=0\,,\label{cond2}\\
&\langle v,s_k\rangle_{L^2}=0\,,\quad\forall k\in \mathbb{N}\,,\label{cond3}\\
&\langle v,c_k\rangle_{L^2}=0\,,\quad\forall k\in \mathbb{N}\,.\label{cond4}
\end{align}
As $c_0$, and all the $s_k\,,c_k$ are proportional to $C_0$, $S_k$ and $C_k$ [which, themselves, are an orthonormal basis of $L^2(I)$)], the conditions \eqref{cond2}, \eqref{cond3} and \eqref{cond4}, alone, imply $v=0$.

Another interesting fact to note is that the image of $T$ is dense in $H^1(I)$. This is so because a subspace $S$ of a Hilbert space $\mathcal{H}$ is dense in  $\mathcal{H}$ if and only if its orthogonal $S^\perp=\{0\}$. Now, if $w\in H^1(I)$ satisfies $\langle T[v],w\rangle_{H^1}=0$ for all $v\in H^1(I)$, then $\langle v,w\rangle_{L^2}=0$, also for all $v\in H^1(I)$, so, in particular we have
\begin{align*}
&\langle w,c_0\rangle=0\,,\\
&\langle w,c_k\rangle=0\,,k\in\mathbb{N}\,,\\
&\langle w,s_k\rangle=0\,, k\in\mathbb{N}\,,
\end{align*}
As $c_0$, $c_k$ and $s_k$ are proportional to $C_0$, $C_k$ and $S_k$, these vectors constitute an orthonormal basis of $L^2(I)$, and $w\in L^2(I)$ we see that, necessarily, $w=0$.

As a final comment, it is important to notice that \eqref{Tsegunda} provides a simple way to invert $T$ over its image as $v=T[v]-a^2T[v]''$. As a consequence of this, no linear combination of $\exp(x/a)$ and $\exp(-x/a)$ is in the image of $T$ (which proves, again, that $T$ is not onto). 

The following expressions for solutions to the field equations \eqref{field_eqs} of the form provided by d'Alembert's formula are useful to understand the conditions that the scalar field and its derivatives satisfy at the boundary of the interval $[-\ell,\ell]$:

\begin{align}\label{derivatives_odd}
\begin{split}
\hspace*{-.2cm}\partial_x^{2n-1}\Phi(t)&=\frac{1}{a^{2}\sinh\left(\frac{2\ell}{a}\right)}\left(\int_{-\ell}^\ell\!\!\partial_t^{2n-2}\big(\Phi(t,\xi)-a^2\ddot{\Phi}(t,\xi)\big)\sinh\left(\frac{\xi}{a}\right)\mathrm{d}\xi\right)\cosh\left(\frac{x}{a}\right)\\
&+\frac{1}{2a^{2}\sinh\left(\frac{\ell}{a}\right)}\int_{-\ell}^x\!\!\partial_t^{2n-2}\big(\Phi(t,\xi)-a^2\ddot{\Phi}(t,\xi)\big)\sinh\left(\frac{x-\xi-\ell}{a}\right)\mathrm{d}\xi\\
&+\frac{1}{2a^{2}\sinh\left(\frac{\ell}{a}\right)}\int_x^{\ell}\!\!\partial_t^{2n-2}\big(\Phi(t,\xi)-a^2\ddot{\Phi}(t,\xi)\big)\sinh\left(\frac{x-\xi+\ell}{a}\right)\mathrm{d}\xi\,,
\end{split}
\end{align}
\begin{align}\label{derivatives_even}
\begin{split}
\hspace*{-.2cm}\partial_x^{2n}\Phi(t)&=\frac{1}{a\sinh\left(\frac{2\ell}{a}\right)}\left(\int_{-\ell}^\ell\!\!\partial_t^{2n}\big(\Phi(t,\xi)-a^2\ddot{\Phi}(t,\xi)\big)\sinh\left(\frac{\xi}{a}\right)\mathrm{d}\xi\right)\sinh\left(\frac{x}{a}\right)\\
&+\frac{1}{2a\sinh\left(\frac{\ell}{a}\right)}\int_{-\ell}^x\!\!\partial_t^{2n}\big(\Phi(t,\xi)-a^2\ddot{\Phi}(t,\xi)\big)\cosh\left(\frac{x-\xi-\ell}{a}\right)\mathrm{d}\xi\\
&+\frac{1}{2a\sinh\left(\frac{\ell}{a}\right)}\int_{x}^\ell\!\!\partial_t^{2n}\big(\Phi(t,\xi)-a^2\ddot{\Phi}(t,\xi)\big)\cosh\left(\frac{x-\xi+\ell}{a}\right)\mathrm{d}\xi\,,
\end{split}
\end{align}
[here $n\in\mathbb{N}$ and $\Phi(t,\xi):=\Phi(t)(\xi)$]. Completely analogous expressions, which can be obtained by differentiating \eqref{derivatives_odd} and \eqref{derivatives_even} with respect to $t$, hold for $\partial_x^{2n-1}\dot{\Phi}(t)$ and $\partial_x^{2n}\dot{\Phi}(t)$. They can all be proved by starting from
\[
\Phi(t)=T\left[\Phi(t)-a^2\ddot{\Phi}(t)\right]\,,
\]
and using the explicit form of $T$ given by \eqref{def_T}.

%
%
\section{Solutions to d'Alembert's equation satisfy the field equations}\label{appendix_computations}

As stated at the end of Section \ref{sec_Lag} the solutions to the wave equation \eqref{wave_eq} with the form \eqref{DAlembert} satisfy the integro-differential field equations \eqref{field_eqs} if the functions $F$ and $G$ are smooth. This can be proved by plugging  \eqref{DAlembert} into \eqref{field_eqs}. In order to do this it is convenient to consider the terms involving $F$ and $G$ separately.

When the terms with $F$ are introduced into \eqref{field_eqs} one gets expressions involving integrals of $F$ added to integrals involving second derivatives of $F$. Integrating the latter by parts twice and gathering the boundary terms one arrives at
\begin{align*}
&-\frac{a}{4\cosh\left(\frac{\ell}{a}\right)}\big(F'(\ell+t)+F'(\ell-t)+F'(-\ell+t)+F'(-\ell-t)\big)\sinh\left(\frac{x}{a}\right)\\
&+\frac{a}{4\sinh\left(\frac{\ell}{a}\right)}\big(F'(-\ell+t)+F'(-\ell-t)-F'(\ell+t)-F'(\ell-t)\big)\cosh\left(\frac{x}{a}\right)\\
&+\frac{1}{2}\big(F(x+t)+F(x-t)\big)\,.
\end{align*}
Taking now into account that, according to the way the function $F$ is built by extending the initial data $\Phi_0$ to $\mathbb{R}$, we have $F'(x)+F'(2\ell-x)=0$ and $F'(x-2\ell)+F'(-x)=0$ for all $x\in[-\ell,\ell]$, the previous expression simplifies to
\[
\frac{1}{2}\big(F(x+t)+F(x-t)\big)\,.
\]

The computation of the  $G$ dependent terms involves double integrals and integrals where first order derivatives of $G$ appear. The way to proceed is to exchange the order of integration in the double integrals and integrate by parts in the integrals with $G'$. This way we get
\begin{align*}
&-\frac{a}{4\cosh\left(\frac{\ell}{a}\right)}\big(G(\ell+t)-G(\ell-t)+G(-\ell+t)-G(-\ell-t)\big)\sinh\left(\frac{x}{a}\right)\\
&+\frac{a}{4\sinh\left(\frac{\ell}{a}\right)}\big(G(-\ell+t)-G(-\ell-t)+G(\ell-t)-G(\ell+t)\big)\cosh\left(\frac{x}{a}\right)\\
&+\frac{1}{2}\int_{x-t}^{x+t}G(\tau)\mathrm{d}\tau\,.
\end{align*}
After doing this, and taking into account that for $x\in[-\ell,\ell]$ the function $G$ satisfies $G(2\ell-x)-G(x)=0$ and $G(-x)-G(x-2\ell)=0$, one finds
\[
\frac{1}{2}\int_{x-t}^{x+t}G(\tau)\mathrm{d}\tau\,.
\]

\end{appendices}

%
%

\end{document}